\documentclass[
    ,final            
  ]
  {aipproc}

\layoutstyle{6x9}

\newcommand{\eg}{e.g., }
\newcommand{\ie}{i.e., }

\newcommand{\Nifs}{$^{56}$Ni}

\newcommand{\KE}{E_{\rm K}}
\def\gsim{\mathrel{\rlap{\lower 4pt \hbox{\hskip 1pt $\sim$}}\raise 1pt
\hbox {$>$}}}
\def\lsim{\mathrel{\rlap{\lower 4pt \hbox{\hskip 1pt $\sim$}}\raise 1pt
\hbox {$<$}}}

\newcommand{\samurai}{\texttt{SAMURAI}}

\def\ion#1#2{{\rm #1}~{\sc #2}}

\begin{document}

\title{Multi-Dimensional Simulations of Radiative Transfer 
in Aspherical Core-Collapse Supernovae}


\author{Masaomi Tanaka}{
  address={Department of Astronomy, Graduate School of Science, 
University of Tokyo, Tokyo, Japan; mtanaka@astron.s.u-tokyo.ac.jp}
}

\author{Keiichi Maeda}{
  address={Institute for the Physics and Mathematics of the Universe, 
University of Tokyo, Kashiwa, Japan}
  ,altaddress={Max-Planck-Institut f\"{u}r Astrophysik, 
Garching bei M\"{u}nchen, Germany}
}

\author{Paolo A. Mazzali}{
  address={Max-Planck-Institut f\"{u}r Astrophysik, 
Garching bei M\"{u}nchen, Germany}
 ,altaddress={Istituto Nazionale di Astrofisica, OATs, Trieste, Italy}
}

\author{Ken'ichi Nomoto}{
  address={Institute for the Physics and Mathematics of the Universe, 
University of Tokyo, Kashiwa, Japan}
  ,altaddress={Department of Astronomy, Graduate School of Science, 
University of Tokyo, Tokyo, Japan; mtanaka@astron.s.u-tokyo.ac.jp} 
}

\begin{abstract}

We study optical radiation of aspherical supernovae (SNe)
and present an approach to verify the asphericity of SNe 
with optical observations of extragalactic SNe.
For this purpose, we have developed
a multi-dimensional Monte-Carlo radiative transfer code, 
\texttt{SAMURAI} \ (SupernovA MUlti-dimensional RAdIative transfer code).
The code can compute the optical light curve and spectra both at 
early phases ($\lsim$ 40 days after the explosion) and late phases
($\sim$ 1 year after the explosion),
based on hydrodynamic and nucleosynthetic models.
We show that all the optical observations of SN 1998bw 
(associated with GRB 980425) are consistent with polar-viewed 
radiation of the aspherical explosion model with kinetic energy
$20 \times 10^{51}$ ergs.
Properties of off-axis hypernovae are also discussed briefly.

\end{abstract}
\maketitle


\section{Introduction}
\label{sec:introduction}

Although the explosion mechanism of core-collapse supernovae (SNe) 
is not well understood,
there is several observational evidence of non-spherical explosion,
obtained by the imaging of {\it very} nearby SNe, 
\eg SN 1987A \citep{wan02} and Galactic supernova remnants \citep{hwa04}.
Even when the imaging is not possible, 
the detection of polarization from extragalactic SNe 
\citep{wan01, kaw02, leo06} suggests that SNe are not spherical.
In addition to these studies, 
spectroscopy of SNe can also give clues of the structure of SN explosion.
For example, emission line profiles in the late time spectra 
($t \gsim$ 1 year, where $t$ is the time after the explosion)
reveals the asphericity of the explosion \citep{maz05}.

It is well established that 
a special class of Type Ic SNe
\footnote{SNe that do not show H, He, and strong Si
absorption in the early time spectra ($t \lsim 40$ days)
are classified as Type Ic \citep{fil97}.} 
are associated with the long gamma-ray bursts (GRBs, see \citep{woo06}
and references therein).
This class of SNe is thought to be highly
energetic, so called hypernovae
(here defined as SNe with ejecta kinetic energy 
$E_{51} = \KE / 10^{51} {\rm ergs} > 10$; \eg \citep{nom06}).
The asphericity of hypernovae is of great interest
related to the nature of GRBs.

Since GRBs are induced by relativistic jets, 
hypernovae are also thought to be aspherical.
However, the large kinetic energy of hypernovae 
is estimated by the analysis under the spherical symmetry.
No realistic multi-dimensional explosion models have been 
verified against the observed early phase spectra.

To study the multi-dimensional nature of the SN explosion
through the various observational facts, 
radiative transfer calculations are required to
connect observables and hydrodynamics models.
In this paper, 
radiative transfer in SN ejecta is solved 
with a multi-dimensional Monte-Carlo radiative transfer code, \samurai
\ (SupernovA MUlti-dimensional RAdIative transfer code),
based on hydrodynamic and nucleosynthetic models of hypernovae.
The results are compared with observations of SN 1998bw, and
implications for off-axis hypernovae are discussed.

\begin{figure}
\includegraphics[scale=0.6, angle=270]{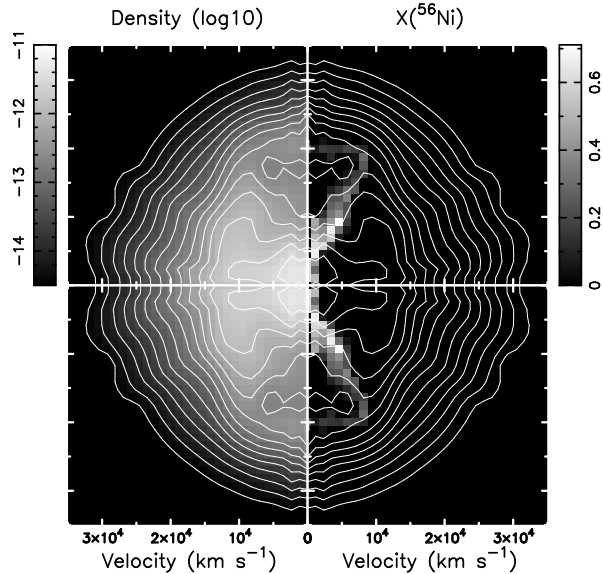}
\caption{
Aspherical explosion model A20 with the kinetic energy
$E_{51} = \KE / 10^{51} {\rm ergs} =20$.
{\it Left}: Density distribution (log ${\rm g \ cm^{-3}}$) at t=10 days.
The contour also shows the density. 
{\it Right}: Mass fraction of \Nifs.
The velocity can be used as spatial coordinate
thanks to the homologous expansion ($r \propto v$).
\label{fig:model}}
\end{figure}

\section{Explosion Models}
\label{sec:models}

We use the results of multi-dimensional hydrodynamic 
and nucleosynthetic calculations for SN 1998bw \citep{mae02} 
as input density and element distributions.
Figure \ref{fig:model} shows the density structure and 
the distribution of \Nifs.
Since the original models used a He star as a progenitor,
we simply replace the abundance of the He layer with that of the C+O layer.
In the hydrodynamic model, energy is deposited aspherically, with 
more energy in the jet direction (z-axis, defined as $\theta = 0^{\circ}$).
As a result, \Nifs\ is preferentially synthesized along this direction 
(right panel of Fig. \ref{fig:model}).
In this paper, an aspherical model with $E_{51} =20$ (A20)
and a spherical model with $E_{51} =50$ (F50) are studied.
They are constructed based on the models with $E_{51}=10$ 
\citep{mae02, mae06a}.

\section{The Numerical Code}
\label{sec:method}

\begin{figure}
\begin{tabular}{cc}
\includegraphics[scale=0.9]{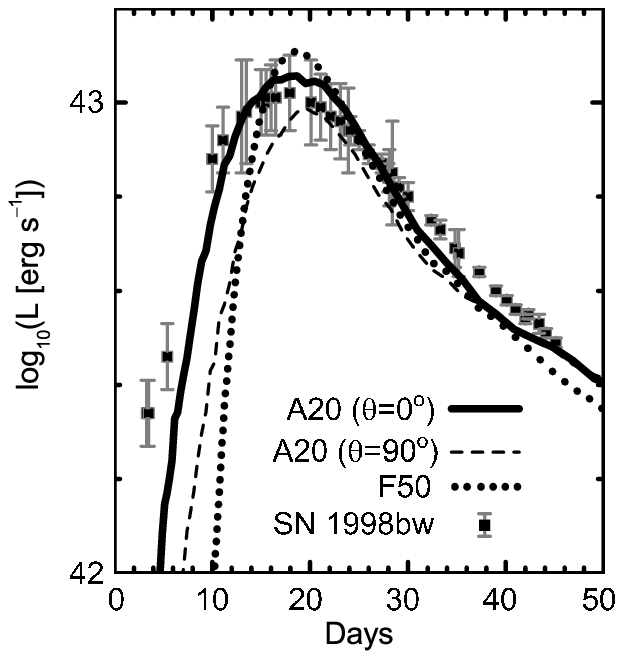}&
\includegraphics[scale=0.9]{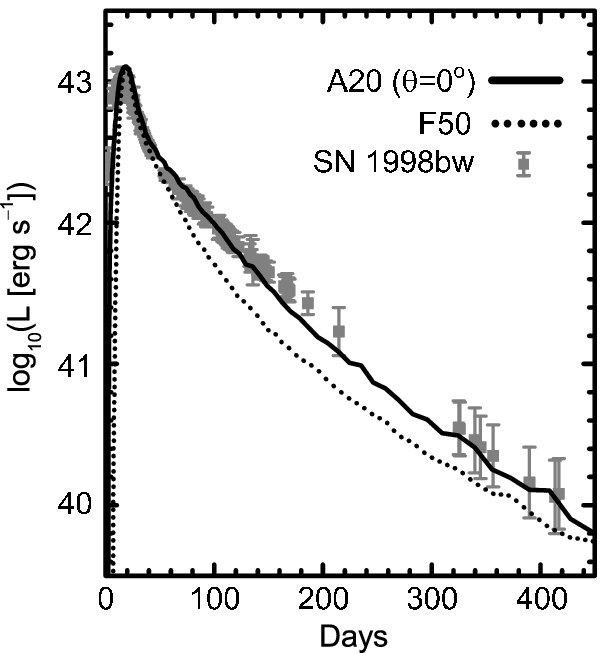}
\end{tabular}
\caption{The LC of SN 1998bw (points) compared with 
the results of simulations.
{\it Left}: The LCs at early phases. The LC of the polar-viewed model 
($\theta = 0^{\circ}$) rises earlier than that of the side-viewed model
($\theta = 90^{\circ}$).
{\it Right}: The LCs including late phases. 
The LC of the spherical model that explained early phase LC and spectra (F50)
fades faster than the observed LC. 
\label{fig:LC}}
\end{figure}

In order to study the detailed properties of the radiation 
from aspherical SNe, 
we have unified a SupernovA MUlti-dimensional RAdIative transfer code \samurai.
\samurai\ is a combination of 3D codes adopting Monte-Carlo 
methods to compute the bolometric light curve (LC) \citep{mae06a,mae06}, and
the spectra of SNe from early \citep{tan06, tan07} to late phases 
\citep{mae06b} 
\footnote{See also \citep{hoe96, hoe99, tho02, kas04, koz05, sim07}
for other multi-dimensional codes.}.

The early phase spectra are calculated as snapshots 
in the optically thin atmosphere, using the results of 
the LC simulation as initial conditions.
A sharply defined photosphere is assumed as an inner boundary for simplicity.
The position of the inner boundary in each direction is 
determined by averaging the positions of the last scattering photon packets 
(see Fig. 3 of \citet{mae06a}).
For the computation of ionization and excitation state
in the atmosphere,
the local physical process same as in the previous 1D code 
\citep{maz93, maz00}.
Line scattering under the Sobolev approximation
and electron scattering are taken into account.
For line scattering, the effect of photon branching is included
as in \citet{luc99}.
In the simulations, 16 elements are included,
\ie H, He, C, N, O, Na, Mg, Si, S, Ti, Cr, Ca, Ti, Fe, Co and Ni.

\section{Light Curves}
\label{sec:LC}

\citet{mae06a} computed bolometric LCs in 3D space.
A common problem in hypernova LCs is  
that a spherical model reproducing the LC and spectra at early phases 
($E_{51} = 50$ for SN 1998bw) declines more rapidly than the observed LC
at $t \gsim 100$ days \citep{nak01, mae03} 
(see model F50 in the right panel of Fig. \ref{fig:LC}).
This problem can be solved by aspherical models with a polar view.
In aspherical models, even with a lower kinetic energy
($E_{51} = 10 - 20$, model A20 in Fig. \ref{fig:LC}) than 
in the spherical case ($E_{51} = 50$),
which allows sufficient trapping of $\gamma$-rays at late times,
the rapid rise of the LC can be reproduced
because of the extended \Nifs\ distribution \citep{mae06a}.

\section{Spectra}
\label{sec:spectra}

\begin{figure}
\begin{tabular}{cc}
\includegraphics[scale=0.48]{f3a.eps}&
\includegraphics[scale=0.95]{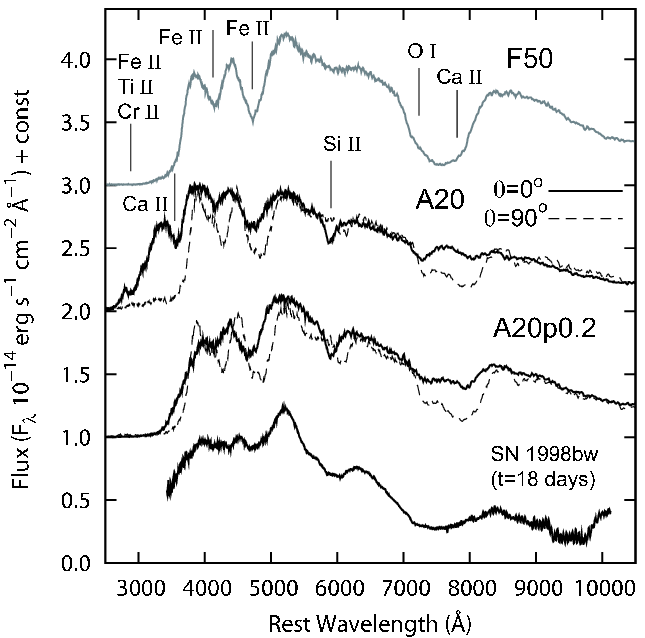}
\end{tabular}
\caption{
{\it Left}: Temperature structure ({\it upper left}), Ca mass fraction
({\it upper right}), ionization fraction of \ion{Ca}{ii} and \ion{Ca}{iii}
({\it lower left} and {\it lower right}, respectively) in 
the SN atmosphere at $t=20$ days.
{\it Right}: The observed spectrum of SN 1998bw at t=18 days 
compared with the synthetic spectra computed with model F50 (spherical), 
A20 (aspherical) and A20p0.2 (aspherical $+$ mixing).
For aspherical models, the solid and dashed lines show the spectra
viewed from polar ($\theta = 0^{\circ}$) 
and equatorial ($\theta = 90^{\circ}$) direction, respectively.
The synthetic spectra are scaled to match the observed spectra 
since the input luminosity is slightly brighter than the observation
(Fig. \ref{fig:LC}, see \citet{tan07} for details).
The synthetic spectra are shifted by 3.0, 2.0, 1.0 $\times 10^{-14}$
from top to bottom.
\label{fig:specd20}}
\end{figure}

The right panel of Figure \ref{fig:specd20} shows the synthetic spectra
at $t=20$ days (around the maximum brightness) 
for models F50 and A20 compared with 
the observed spectrum of SN 1998bw.
In model A20, all the absorption lines 
except for \ion{Si}{ii} $\lambda$6355 are
stronger for larger $\theta$, \ie for a side view.
This is understood by the asphericity of the temperature structure 
in the SN atmosphere.
As shown in the left panel of Figure \ref{fig:specd20}, 
the temperature near the z-axis is higher than in 
the equatorial plane by $\sim 2000$ K ({\it upper left}),
tracing the aspherical distribution of \Nifs.
This makes the ionization degree near the z-axis higher
({\it lower} panels).
As a result, all species that have strong lines,
\ie \ion{O}{i}, \ion{Si}{ii}, \ion{Ca}{ii}, \ion{Ti}{ii}, \ion{Cr}{ii}
and \ion{Fe}{ii}, dominate near the equator but not near the z-axis.

\begin{figure}
\begin{tabular}{cc}
\includegraphics[scale=0.48]{f4a.eps}&
\includegraphics[scale=0.95]{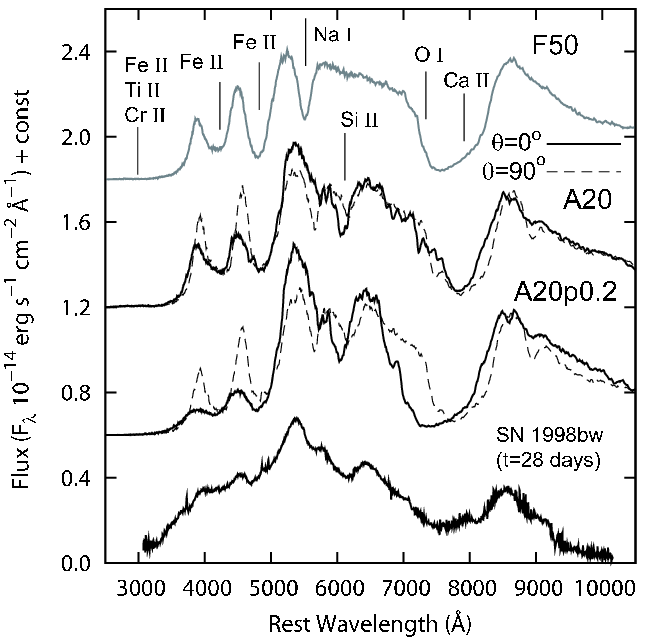}
\end{tabular}
\caption{
Same as Figure \ref{fig:specd20} but at $t=30$ days.
The synthetic spectra are shifted by 1.8, 1.2, 0.6 $\times 10^{-14}$
from top to bottom.
\label{fig:specd30}}
\end{figure}

The right panel of Figure \ref{fig:specd30} shows the synthetic spectra
at $t=30$ days for models F50 and A20.
The emergent spectra of the aspherical model are not 
significantly different for different viewing angles (Fig. \ref{fig:specd30}).
At this epoch, the temperature structure are still 
anisotropic, and consequently, 
the distribution of ionization fractions is also aspherical 
(see the left panel of Fig. \ref{fig:specd30}).
However, since nucleosynthesis occurs entirely near the polar 
direction in the model, and the photosphere at this epoch is
located inside the region where heavy elements are synthesized 
in the explosion,
the suppression of important ions near the z-axis is compensated by the larger 
abundance of the heavy elements
(see $X$(Ca) in Fig. \ref{fig:specd30}).

We compare the polar-viewed spectra of model A20 with the observed 
spectra of SN 1998bw.
At $t=20$ days, the absorptions of \ion{O}{i}, \ion{Ca}{ii}
and \ion{Fe}{ii}/\ion{Fe}{iii} in the model are weaker 
than in SN 1998bw.
At $t=30$ days, the \ion{Ca}{ii} and 
\ion{Fe}{ii} lines become strong, although the 
\ion{O}{i}-\ion{Ca}{ii} absorption at 7000 -- 8000 \AA\ is still narrower than 
in the observed spectrum. 
In the synthetic polar-viewed spectrum,
the peaks around 4000 and 4500 \AA\ are partially suppressed 
by the high velocity absorption by the extended Fe near the jet, 
while they are strong in the side-viewed spectrum. 
The suppression of the peaks is similarly seen in the spectrum of SN 1998bw.

The strengths of the \ion{Ca}{ii} and \ion{Fe}{ii} lines at $t=20$ days 
can be increased if heavy elements synthesized 
in the explosion are mixed to outer layers.
In SN explosions, Rayleigh - Taylor (R-T) instabilities are expected to occur
(see \citet{kif00} for the case of Type Ic SNe),
which could deliver the newly synthesized elements to higher velocities.
In Figures \ref{fig:specd20} and \ref{fig:specd30}, 
synthetic spectra of model A20p0.2 are also shown.
In this model, $20 \%$ of synthesized material is assumed to be mixed 
to the outer layers.
The agreement with the observed spectra becomes better especially at 
$t=20$ days.

\section{Discussion}
\label{sec:discussion}

We have presented the detailed simulations of 
optical radiation with realistic jet-like hypernova models.
The emergent LC and spectra are different for different viewing angles.
The spectral properties are determined by the combination of 
aspherical abundances and anisotropic ionization states.
Although the agreement of the spectra is far from perfect, 
the spectra of the model with mixing are in qualitative agreement 
with those of SN 1998bw.

The LC study shows that 
the kinetic energy of an aspherical model that explains SN 1998bw
is $E_{51} = 20$, which is less than 
that of a well-fitting spherical model ($E_{51} = 50$).
The early phase spectra can also be explained by the model with
$E_{51} = 20$.
However, it should be noted that the higher kinetic energy than the canonical
SNe ($E_{51} \sim 1$) is still required.

The simulations enable us to predict the 
radiation from off-axis hypernovae.
The LC viewed off-axis rises more slowly than that of on-axis,
and its maximum brightness is fainter (Fig. \ref{fig:LC}).
The spectra viewed off-axis show
(1) a slightly lower absorption velocity, 
(2) stronger peaks around 4000 and 4500 \AA\ (narrower absorption of Fe) and
(3) a stronger \ion{Na}{i} $\lambda$5890 line.
However, the spectra still show general appearance of 
``hypernovae'' or ``broad-line supernovae''.
At later phases, off-axis hypernovae would show 
double-peaked [\ion{O}{i}] emission profile \citep{maz05, mae06b}.


\begin{theacknowledgments}
M.T. is supported through the JSPS (Japan Society for the 
Promotion of Science) Research Fellowship for Young Scientists.
\end{theacknowledgments}

\end{document}